\documentclass[%
 reprint,
 amsmath,amssymb,
 aps,
 prb,
 showpacs,
 superscriptaddress
]{revtex4-2}

\usepackage{graphicx}
\usepackage{dcolumn}
\usepackage{bm}
\usepackage{comment}
\usepackage[colorlinks,citecolor=blue,linkcolor=blue,urlcolor=blue]{hyperref}

\begin{document}

\preprint{AAPM/123-QED}

\title[Sample title]{Ab-initio study of magneto-ionic mechanisms in ferromagnet/oxide multilayers}
\author{Adriano Di Pietro$^{1,2}$, Rohit Pachat$^3$, Liza Herrera Diez$^3$, Johannes W. van der Jagt$^{4,5}$, Dafiné Ravelosona$^{3,4}$, Gianfranco Durin$^1$
\smallskip 
\\
$^1$ \textit{Istituto Nazionale di Ricerca Metrologica, Strada delle Cacce 91, 10135, Torino, Italy}
\\
$^2$\textit{Politecnico di Torino, Corso Duca degli Abruzzi 24, 10129, Torino,  Italy}
\\
$^3$\textit{Centre  de  Nanosciences  et  de  Nanotechnologies,  CNRS,  Université  Paris-Saclay,  91120  Palaiseau, France}
\\
$^4$ \textit{Spin-ion Technologies, Boulevard Thomas Gobert 10, 91120,  Paliseau, France }
\\ 
$^5$ \textit{Université Paris Saclay, 3 rue Juliot Curie, 91190, Gif-sur-Yvette, France}
}
\date{\today}

\begin{abstract}
The application of gate voltages in heavy metal/ferromagnet/Oxide multilayer stacks has been identified as one possible candidate to manipulate their anisotropy at will. However, this method has proven to show a wide variety of behaviours in terms of reversibility, depending on the nature of the metal/oxide interface and its degree of oxidation. In order to shed light on the microscopic mechanism governing the complex magneto-ionic behaviour in $\text{Ta/CoFeB/}\text{HfO}_2$, we perform ab-initio simulations on various setups comprising $\text{Fe/O, Fe/HfO}_2$ interfaces with different oxygen atom interfacial geometries. After the determination of the more stable interfacial configurations, we calculate the magnetic anisotropy energy on the different unit cell configurations and formulate a possible mechanism that well describes the recent experimental observations in $\text{Ta/CoFeB/}\text{HfO}_2$.
\end{abstract}

\maketitle

\section{\label{sec:level1} Introduction}

The  ever-increasing demand for memory storage in the modern IT industry has made the need for new energy efficient storage alternatives all the more important.
Voltage control of magnetic anisotropy (VCMA) \cite{One2021} has gained scientific interest as one of the prime candidates to develop ultra-low energy memory storage devices \cite{SUW-19,DIE-17,Hallal2013,Belabbes2016} and is usually studied in 2 different variants: the first one aims at modifying magnetic properties of thin films by pure charge accumulation/depletion effects induced by voltage application \cite{IBR-16,HUA-13,Hallal2013,Niranjan2010}. The second variant makes use of voltage induced ionic motion in heavy metal (HM)/ferromagnet (FM)/Oxide (Ox) thin film multilayers to carefully tune the oxygen/ferromagnet chemical and electrostatic interaction, enabling the control of magnetic anisotropy \cite{Monso2002, Bauer2015, DIE-17}. The main advantage of ionic manipulation in comparison to pure charge accumulation/depletion techniques is the non-volatility of the magnetization switching, while the trade off is a more complex reversibility mechanism combining ion mobility and chemical composition at the FM/Ox interface. The work of Fassatoui et al. \cite{Fassatoui2020} clearly shows how the application of gate voltages causes a reversible magnetization switching in Pt/Co/AlOx as well as in 
Pt/Co/TbOx. At the same time, applying a gate voltage in Pt/Co/MgOx has the effect of irreversibly pushing the anisotropy easy axis out-of-plane. This discrepancy has been attributed to the result of the different character of the ionic mobility of the oxides: TbOx and AlOx have the common property of being oxides with a predominantly oxygen based ionic mobility \cite{Fassatoui2020,Flynn1962}, while MgOx is known to have Mg as the principal ionic carrier under the application of gate voltages \cite{Wu2017}. 
CoFeB/Oxide multilayer structures are of great technological interest as they have shown promise for the design of non-volatile, high-density memory storage devices thanks to their high tunneling mangeto-resistance (TMR), low damping and perpendicular magnetic anisotropy (PMA) \cite{Liu2014, Ikeda2010, Wang2020}.
A recent work from Pachat et. al \cite{Pachat} highlighted a more complex magneto-ionic behavior in Ta/CoFeB/HfO$_2$ multilayers. The application of a gate voltage to the as-grown material with in-plane anisotropy (IPA) initially causes a non-volatile, irreversible spin-reorientation transition (SRT) to a perpendicular anisotropy state (PMA). Further application of the gate voltage causes the transition to a fully reversible regime. This 2-step process is in contrast with the picture presented in \cite{Fassatoui2020} because ionic mobility in HfO$_2$ is attributed to oxygen \cite{Schie2017a}. 
 To formulate a hypothesis on the mechanism governing these different reversibility behaviors in Ta/CoFeB/HfO$_2$, we perform ab-initio simulations using density functional theory (DFT) to determine the structural and magnetic properties of two FM/Oxide interfaces. The work is structured as follows: In Sec.~\ref{sections:Methods}, we provide the computational details of our simulations and a brief overview of the theoretical framework used to describe magnetic anisotropy in FM/Oxide interfaces. In Sec.~\ref{section:Results} we proceed and analyze the structural properties of two different FM/Oxide interfaces; each one displaying either interstitial or frontal oxygen positioning (see Fig.~\ref{fig:DFT_Cells} and Fig.~\ref{fig:HfO_2/Fe_Ox_mixed}-(c)). After having determined the optimal oxygen configurations of these setups, we proceed and compute the magnetic anisotropy energy of the Fe/HfO$_2$ unit cells. We also explore the role of ionic mobility in determining the magnetic anisotropy properties of Fe/HfO$_2$ interfaces and  highlight how the energy costs involved in  ionic mobility is different depending on the site occupied. In Sec.~\ref{discussion}, we discuss the results and compare them to the experimental data \cite{Pachat} and theoretical predictions \cite{HUA-13,Liang2014} in order to formulate a hypothesis for the appearance of different magneto-ionic regimes in CoFeB/HfO$_2$ multilayers. Finally, in Sec.~\ref{Conclusion} we provide a summary of the findings and outline new possible systems to analyze to further probe our hypothesis. 

\section{\label{sections:Methods}Methods and computational details}
\subsection{\label{Structural relaxation}Structural relaxations}
\par We perform structural relaxation using density functional theory and applying the full-potential linearized augmented plane wave (FLAPW) method \cite{AND-75}, as implemented in the FLEUR code \cite{FLEUR}. In particular, we rely on the generalized gradient approximation for the exchange-correlation potential as implemented by the Perdew–Burke-Ernzerhof (PBE) functional \cite{1996JChPh.105.9982P}.
Since the simulation of amorphous systems such as CoFeB is beyond the capabilities of ab-initio methods, we reduced the ferromagnetic component of the system to the Fe atoms only. This approximation is justified on the basis of the composition of the Co$_{20}$Fe$_{60}$B$_{20}$/HfO$_2$ stacks studied in the literature, which are iron rich \cite{Pachat}. Furthermore, this approximation for CoFeB in ab-initio simulations is commonly used in the literature \cite{LIN-20}.
We designed 5 different unit cells comprising: an Fe/O interface (structures (I) and (II) in Fig.~\ref{fig:DFT_Cells}) composed of 5 magnetic layers (ML) of Fe sandwiched between 2 mono-atomic layers of oxygen on each side, an Fe/HfO$_2$ system (structures (III) and (IV) in Fig.~\ref{fig:DFT_Cells}) composed of 5 ML of Fe atoms sandwiched between 2 ML of HfO$_2$ on each side. Finally, to account for oxygen coming from atmospheric interaction with the sample, we designed an Fe/HfO$_2$ system displaying both frontal and interstitial oxygen atoms \cite{YAN-11} . This system is composed of 5 ML of Fe atoms sandwiched between 2 ML of HfO$_2$ on each side with an additional O-layer located in the interstitial site of 2 Fe atoms (Fig.~\ref{fig:HfO_2/Fe_Ox_mixed}-(c)). We refer to this kind of system as "mixed interface" throughout the work. The in-plane lattice constant of the system is fixed to the value of BCC Fe, i.e $a_{Fe} = 2.87 $ \AA . Before performing structural relaxation, we selected an energy cutoff value for the plane wave basis of $275.5$ eV and a k-point mesh of dimensions $10 \times 10 \times 1$. These parameters allow us to obtain self consistent energies converged to at least $0.009$ eV/Atom. To determine the optimal interfacial configurations of oxygen atoms, structural relaxation in shape and volume is performed up to  $0.001$ eV/\AA. 
\subsection{\label{section:MAE} Magnetic anisotropy energy}
The ab-initio calculation of the magnetic anisotropy energy \cite{Strange1991} is performed according to the following procedure \cite{Zimmermann2019,YAN-11,IBR-16}: after having determined the more stable interfacial geometries via the structural relaxation outlined in \ref{Structural relaxation}, spin orbit coupling (SOC) is included in the Kohn-Sham Hamiltonian and the magnetic anisotropy energy (MAE) is computed by comparing sums of one-electron energies ("Ev-sum" in the MAE plots in Figs. \ref{fig:HfO_2/Fe_MAE},\ref{fig:HfO_2/Fe_Ox_mixed},\ref{fig:HfO_2/Fe_mixed_energy_cost}-(b)) via the magnetic force theorem \cite{Weinert1985,Wang1996}. This method is widely used and has been validated for transition metal interfaces \cite{Bonski2009}. The additional broken symmetries emerging from the inclusion of magnetic moments in the unit cell requires an increase in the k-point mesh to $12 \times 12 \times 1$. Both with and without spin-orbit coupling, the energy cutoff is kept at $275.5$ eV. With these parameters, we are able to obtain self consistent energies converged to at least $0.009$ eV/Atom. 
\subsection{Theoretical background}
The origins of oxygen enabled anisotropy manipulation in FM/Oxide interfaces have been discussed extensively in the literature \cite{Liang2014,YAN-11}. The underlying theoretical frameworks have been developed by Bruno \cite{Bruno1989} and Van Der Laan \cite{VanDerLaan1998}, which successfully linked the appearance of a finite orbital magnetic moment anisotropy to an anisotropy contribution in the energy by means of spin-orbit coupling. These theoretical frameworks predict a heavy dependence of the anisotropy energy on the exact shape and hybridization of the 3d orbitals. Oxygen atoms, when posed at the right distance, can have dramatic effects in the disruption of the otherwise almost anisotropic magnetic moment distribution of the 3d orbitals of transition metal ferromagnets \cite{YAN-11}. In particular, the 3d-electrons tend to hybridize very effectively  with the oxygen $2 p_z$ orbitals. Therefore, depending on the relative position of Fe and O, different 3d-orbitals are going to be more or less involved in the hybridization. If, for instance, we imagine an oxygen atom sitting on top of an Fe atom, the atomic orbitals that are more likely to hybridize have OOP symmetry \cite{Butler2008} ($3d_{z^2} , 3d_{xz} , 3d_{yz}$). This results in an under compensation of orbital magnetic moment coming from electrons in orbitals with IP symmetry ($3 d_{x^2 - y^2} , 3d_{xy}$), which therefore shift the anisotropy easy axis out of the plane \cite{YAN-11} according to the relation

\begin{equation}
\Delta E_{SO} = \xi_{SO} \frac{\Delta \mu}{ 4 \mu_B}
\end{equation}

where $\Delta E_{SO} $ represents the anisotropy energy , $\Delta \mu$ the orbital moment anisotropy and $\xi_{SO}$ the material dependent spin-orbit coupling constant. If on the other hand, the oxygen atom is sitting in the same plane of the Fe atom, hybridization is going to involve orbitals with IP symmetry \{$3 d_{x^2 - y^2} , 3d_{xy}$\} and the orbital magnetic moment under compensation of the \{$3d_{z^2} , 3d_{xz} , 3d_{yz}$\} orbitals will turn the trend and shift the easy axis in-plane. Despite the thinness of the FM layers, one should always consider that the contributions to magnetic anisotropy are not limited to the first layer, but in fact often involve the second layer and possibly beyond (as in the case of Fe/MgO MTJs \cite{IBR-16}).
In addition, the appearance of magnetic anisotropy is still constrained by the symmetries of the crystal field that is coupled to the spin of the electrons via SOC. This implies that different lattice geometries have different angular dependencies of the anisotropy energy \cite{Bruno1989}. BCC structures (which are the ones we are concerned with) are predicted to have the usual relation
\begin{equation}
\Delta E_{SO} = a + b \sin^2{\theta} \label{MAE_fitting}
\end{equation}
where $\theta$ is the spin quantization axis angle with respect to  the $\hat{z}-$axis and $a,b$ represent material dependent constants of anisotropy. An in-plane magnetic anisotropy (IPA) correspond to the minimum of $\Delta E_{SO}$ for $\theta = \frac{\pi}{2}$, while perpendicular magnetic anisotropy (PMA) correspond to the minimum of $\Delta E_{SO}$ for $\theta = 0$.
This is the fitting function that we are going to use in all our MAE calculations presented in Section \ref{section:Results}. 
The advantage of the perturbational approach is that it does not require the comparison of self-consistent energies for the determination of the fitting parameters $a$ and $b$ in \eqref{MAE_fitting}, but rather allows for the use of the magnetic force theorem \cite{Weinert1985,Wang1996} outlined in \ref{section:MAE}.

\begin{figure}
  \centering
  \includegraphics[scale=0.2]{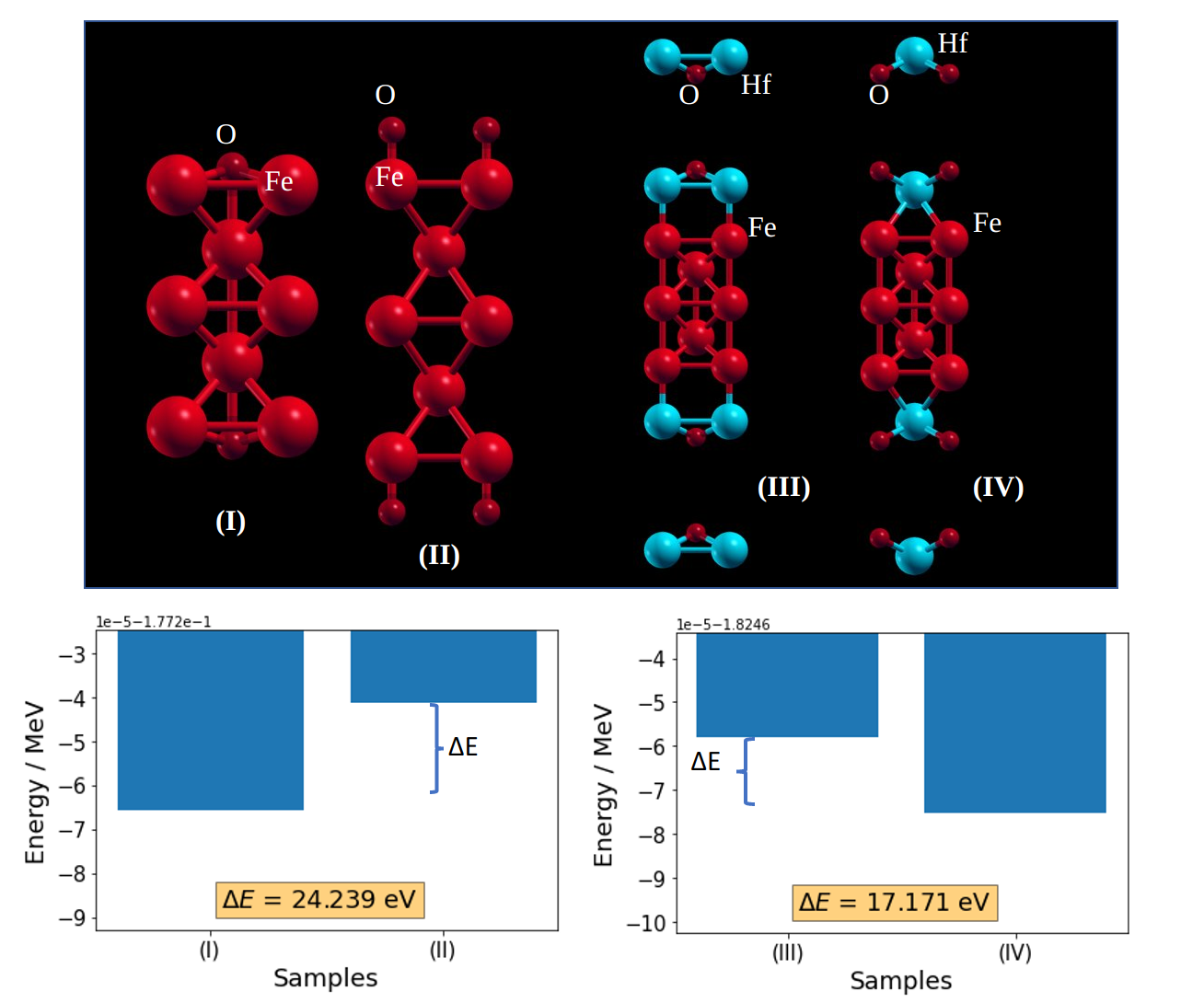}
  \caption{Comparison of the different interfacial configurations in different samples. Unit cells (I) and (II) represent Fe/O interfaces while unit cells (III) and (IV) represent the Fe/HfO$_2$ interfaces. The bar plots represent the ground state energies and highlight the more stable configurations. The yellow highlighted number represents the absolute value of the energy difference between the 2 relaxed structures of each type.}
    \label{fig:DFT_Cells}
\end{figure}

\section{\label{section:Results}Results}
\subsection{Pure interfaces}
\begin{figure*}
  \centering
  \includegraphics[height=9cm]{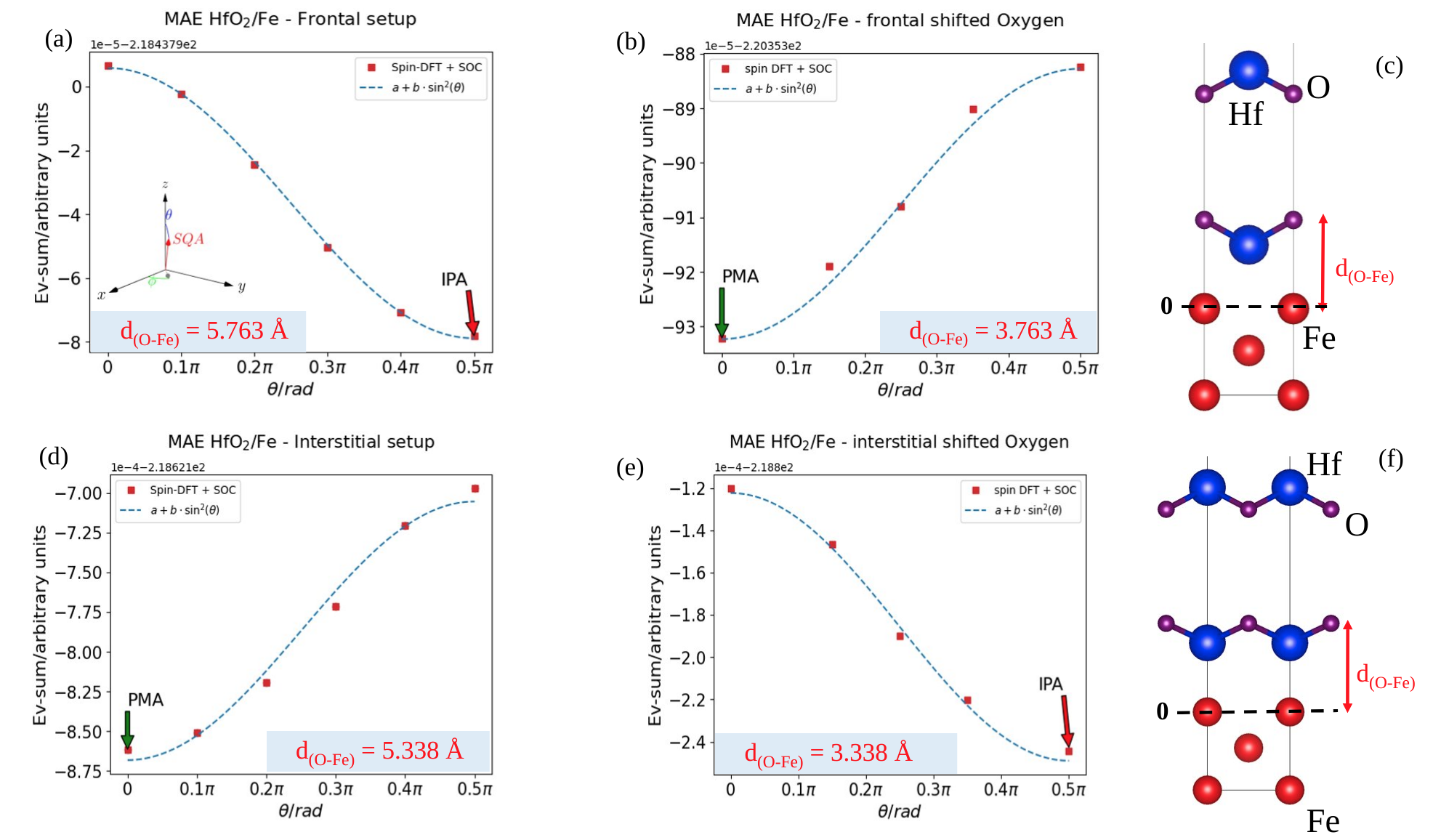}
  \caption{Magnetic anisotropy energy (MAE) comparison of the pure Fe/HfO$_2$ interface (structures (III) and (IV)). The $\theta$ angle of the spin quantization axis (SQA) from eq.\eqref{MAE_fitting} is shown in the inset of (a). MAE for the ground state of the Fe/HfO$_2$ unit cell with frontal Fe-O distance of (a) $5.763 $ \AA \, (equilibrium) and (b) $3.763 $ \AA \, (shifted). (c) Fe/HfO$_2$ unit cell with frontal oxygen positioning. d$_{(O-Fe)}$ represents the distance of the interstitial oxygen species from the Fe surface respectively (marked with the dashed line). Ground state of the Fe/HfO$_2$ unit cell with interstitial Fe-O distance of (d) $5.338$ \AA \, (equilibrium) and (e) $3.338$ \AA \, (shifted). (f) Fe/HfO$_2$ unit cell with interstitial oxygen positioning. d$_{(O-Fe)}$ represents the distance of the interstitial oxygen species from the Fe surface respectively (marked with the dashed line).}
  \label{fig:HfO_2/Fe_MAE}
\end{figure*}

As a first step, we perform structural relaxation on 6 different unit cells where we suppose no interaction with atmospheric oxygen has taken effect (we refer to these as "pure interfaces"). As can be seen by the plotted energies (Fig.~\ref{fig:DFT_Cells}), the optimal oxygen configurations change depending on the system considered: a pure Fe/O interface (structures (I) and (II) of (Fig.~\ref{fig:DFT_Cells})) favours oxygen atoms to be located in the interstitial site with respect to   Fe atoms. This preference appears to be reverted in the case of HfO$_2$, where a frontal positioning of oxygen atoms with respect to  to the Fe atoms seems to be strongly favoured. After having determined the optimal configurations for the different unit cells, we proceed and include spin-orbit coupling to compute their magnetic anisotropy energy. In this case we focus specifically on the Fe/HfO$_2$ interface. By observing Fig.~\ref{fig:HfO_2/Fe_MAE} we notice how the more stable frontally aligned oxygen setup, (structure (IV) of (Fig.~\ref{fig:DFT_Cells})) displays IP magnetic anisotropy (Fig.~\ref{fig:HfO_2/Fe_MAE}-(a)) while an interstitial oxygen configuration (structure (III) of (Fig.~\ref{fig:DFT_Cells})) yields PMA (Fig.~\ref{fig:HfO_2/Fe_MAE}-(d)). At this point, we introduce ionic mobility on the interfacial oxygen atoms and recalculate the magnetic anisotropy. We notice how shifting the frontal oxygen atom in order to reduce the Fe-O distance by $\approx 2$ \AA,  we are able to achieve PMA (Fig.~\ref{fig:HfO_2/Fe_MAE}-(b)). As a side note, we remark how shifting the interstitial oxygen atom of structure (III) in Fig.~\ref{fig:DFT_Cells} has the effect of recovering IP magnetic anisotropy (Fig.~\ref{fig:HfO_2/Fe_MAE}-(e)). 

\begin{figure*}
  \centering
  \includegraphics[height=8.5cm]{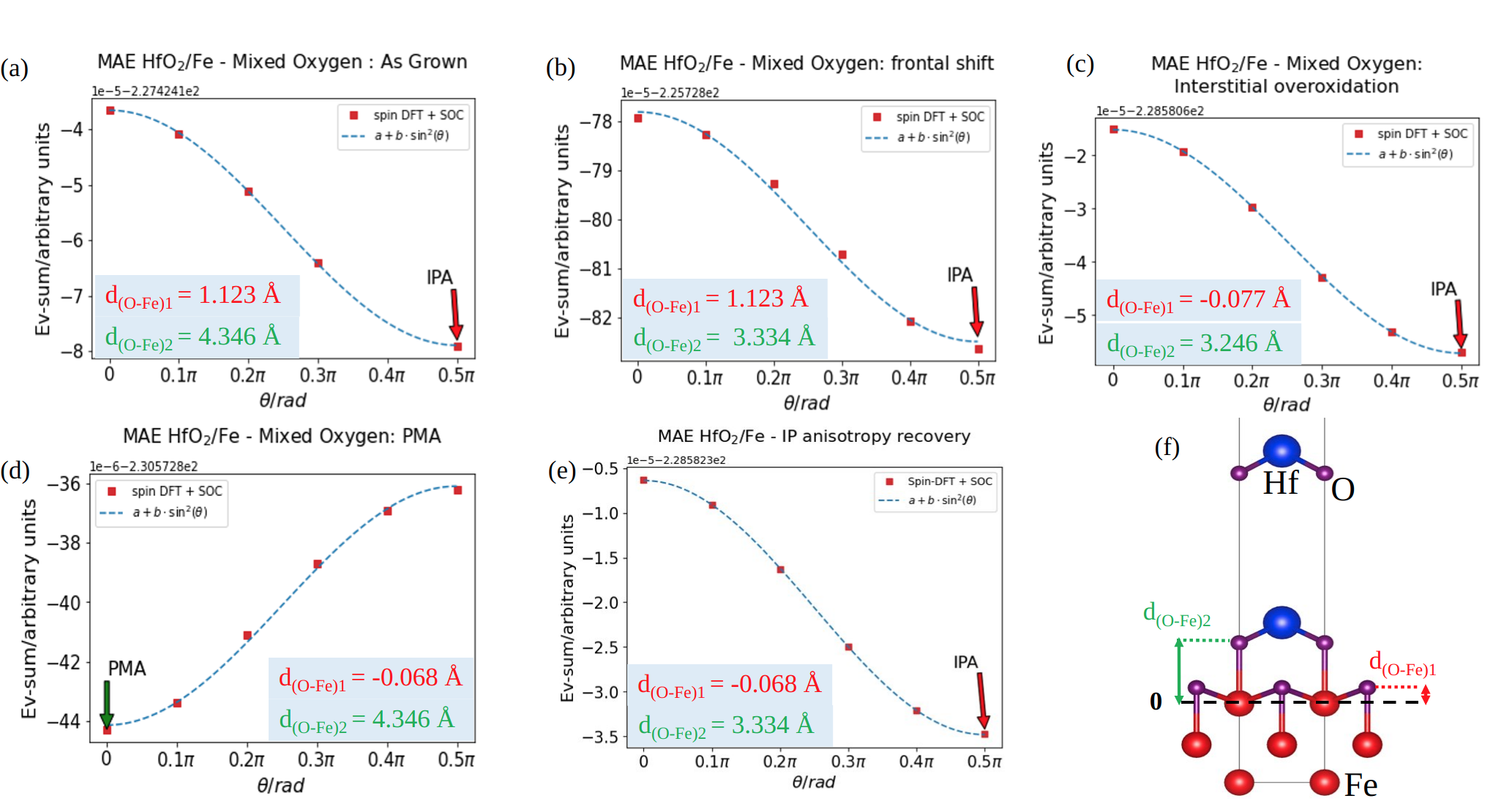}
  \caption{Magnetic anisotropy energy (MAE) ($\theta$ from eq.\eqref{MAE_fitting}) in the Fe/HfO$_2$ setup shown with shifted oxygen species at the interface.  (a) MAE of the mixed surface in its ground state. (b) MAE of the mixed surface with a frontal Fe-O distance of $3.334$ \AA \,. (c) MAE of the mixed surface in with a frontal Fe-O distance of $3.246$ \AA \, and an interstitial Fe-O distance of $-0.077$ \AA. (d) MAE of the mixed surface in with a frontal Fe-O distance of $4.346$ \AA \, and an interstitial Fe-O distance of $-0.068$ \AA. (e) MAE of the mixed surface in with a frontal Fe-O distance of $3.334$ \AA \, and an interstitial Fe-O distance of $-0.068$ \AA. (f) Side view of the Fe/HfO$_2$ unit cell with the mixed setup. The d$_{\text{(O-Fe)}1}$ and d$_{\text{(O-Fe)}2}$ denote the distance of the frontal and interstitial oxygen species from the Fe surface respectively (marked with the dashed line).}
  \label{fig:HfO_2/Fe_Ox_mixed}
\end{figure*}
\begin{figure*}
  \centering
  \includegraphics[width=0.8\textwidth]{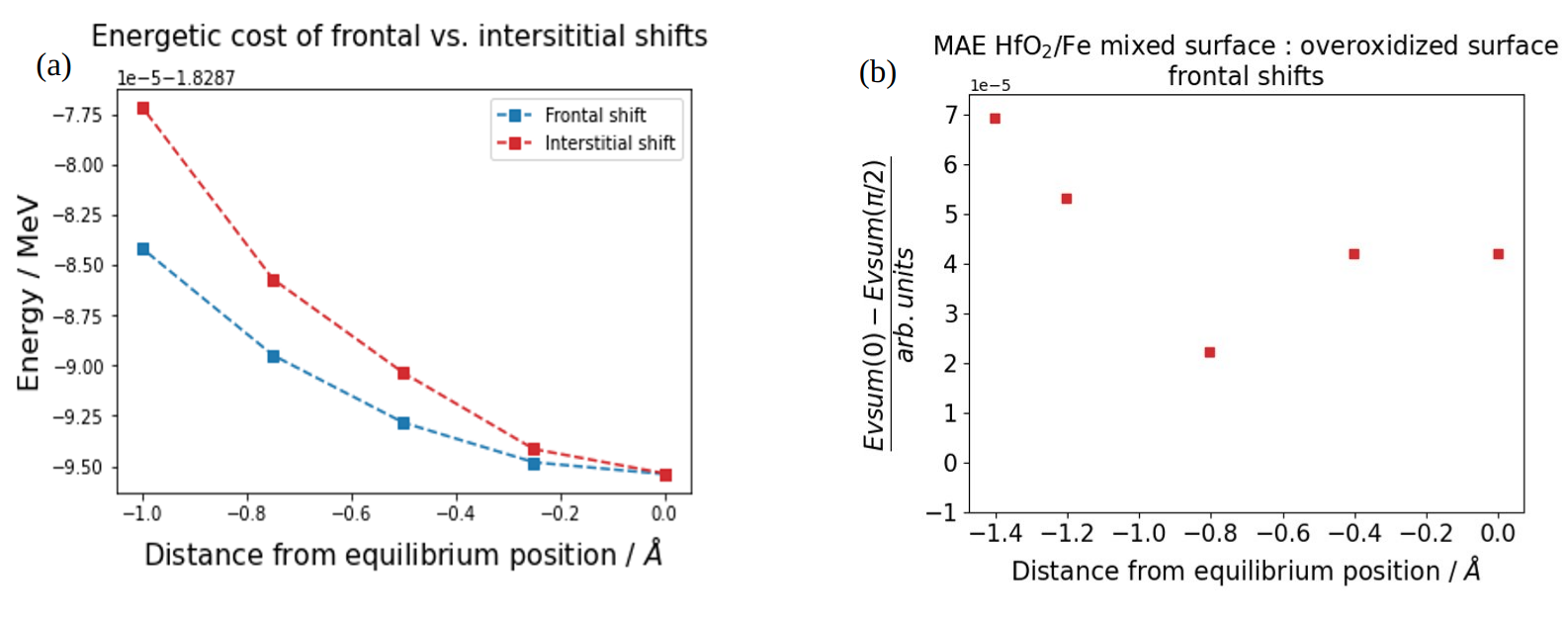}
  \caption{(a) Energetic cost of shifting interstitial and frontal oxygen atoms in the mixed interface setup displayed in panel (f) of Fig.~\ref{fig:HfO_2/Fe_Ox_mixed}. The starting positions are $d_{(O-Fe)1} = 1.123 $ \AA \, for the interstitial oxygen atom and $d_{(O-Fe)2} = 4.436 $ \AA \, for the frontal oxygen atoms (i.e. the setup of panel (a) in Fig.~\ref{fig:HfO_2/Fe_Ox_mixed}) (b) Eigenvalue sum of the system with SQA point OOP - Eigenvalue sum of the system with SQA point IP at different frontal oxygen positions $d_{(O-Fe)2}$ represented in panel (f) of Fig.~\ref{fig:HfO_2/Fe_Ox_mixed}. The starting positions are $d_{(O-Fe)1} = -0.0077 $ \AA \, for the interstitial oxygen atom and $d_{(O-Fe)2} = 3.236 $ \AA \, for the frontal oxygen atoms (i.e the setup of panel (c) in Fig.~\ref{fig:HfO_2/Fe_Ox_mixed}). Positive values on the y-axis indicate that the system has IP magnetic anisotropy.}
  \label{fig:HfO_2/Fe_mixed_energy_cost}
\end{figure*}

\subsection{Mixed Fe/HfO$_2$ interface}

As mentioned in Sec.~\ref{sections:Methods}, we believe the actual experimental setup is not presenting a pure interface. To account for the presence of additional mobile oxygen impurities, we designed an Fe/HfO$_2$ unit cell displaying both interstitial and frontal oxygen alignment at the interface (see Fig.~\ref{fig:HfO_2/Fe_Ox_mixed}-(d)). By analyzing the effect on the magnetic anisotropy of different oxygen species mobility, we expect to be able to paint a plausible picture for the microscopic mechanism governing magneto-ionic regimes in an experimental, more disordered scenario. As can be seen in Fig.~\ref{fig:HfO_2/Fe_Ox_mixed}-(a), the relaxed structure with the both frontal and interstitial oxygen displays IP magnetic anisotropy. If we shift the frontal oxygens by a nominal distance of $1$ \AA  ,\, we notice no effect on magnetic anisotropy (Fig.~\ref{fig:HfO_2/Fe_Ox_mixed}-(b)). If on the other hand, we shift the interstitial oxygen atom by the same amount, we notice how the system displays PMA (Fig.~\ref{fig:HfO_2/Fe_Ox_mixed}-(d)). Despite having checked the effect of oxygen shifts on anisotropy manipulation in this mixed surface setup, we expect the ionic mobility behavior of these two oxygen species to be different given their different environment. By observing Fig.~\ref{fig:HfO_2/Fe_mixed_energy_cost}-(a), we can see that the energy cost of a frontal shift is lower than the energy cost of an interstitial shift, in accordance with our expectations. The energy costs are obtained by comparing the energy of identical unit cells that differ only by the position of the frontal/interstitial oxygen atom \cite{Kyritsakis2019}. If we proceed and further reduce the distance of frontal oxygens from the Fe surface, we notice how the PMA of Fig.~\ref{fig:HfO_2/Fe_Ox_mixed}-(d) is destroyed (Fig.~\ref{fig:HfO_2/Fe_Ox_mixed}-(c)). From Fig.~\ref{fig:HfO_2/Fe_Ox_mixed}-(e) we can see how reducing the frontal Fe-O distance after PMA is achieved restores IP magnetic anisotropy. We also point out how the disruption of PMA can also be obtained by pushing the interstitial oxygens deeper in the ferromagnetic layer (Fig.~\ref{fig:HfO_2/Fe_Ox_mixed}-(c)) \cite{YAN-11}. From Fig.~\ref{fig:HfO_2/Fe_mixed_energy_cost}-(b) we can however observe that once PMA is destroyed by this type of oxygen incorporation, it cannot be restored by moving the frontal oxygen atoms closer to the surface. This observation is also in agreement with the experiments \cite{Pachat}, where samples that were exposed to a negative gate voltage for long times (for a reference of the field direction, see Fig.~\ref{micro_mech}) did not display any reversibility of the voltage induced SRT, which was attributed to over-oxidation.
\section{\label{discussion}Discussion}

\begin{figure*}
  \centering
\includegraphics[height=6.3cm]{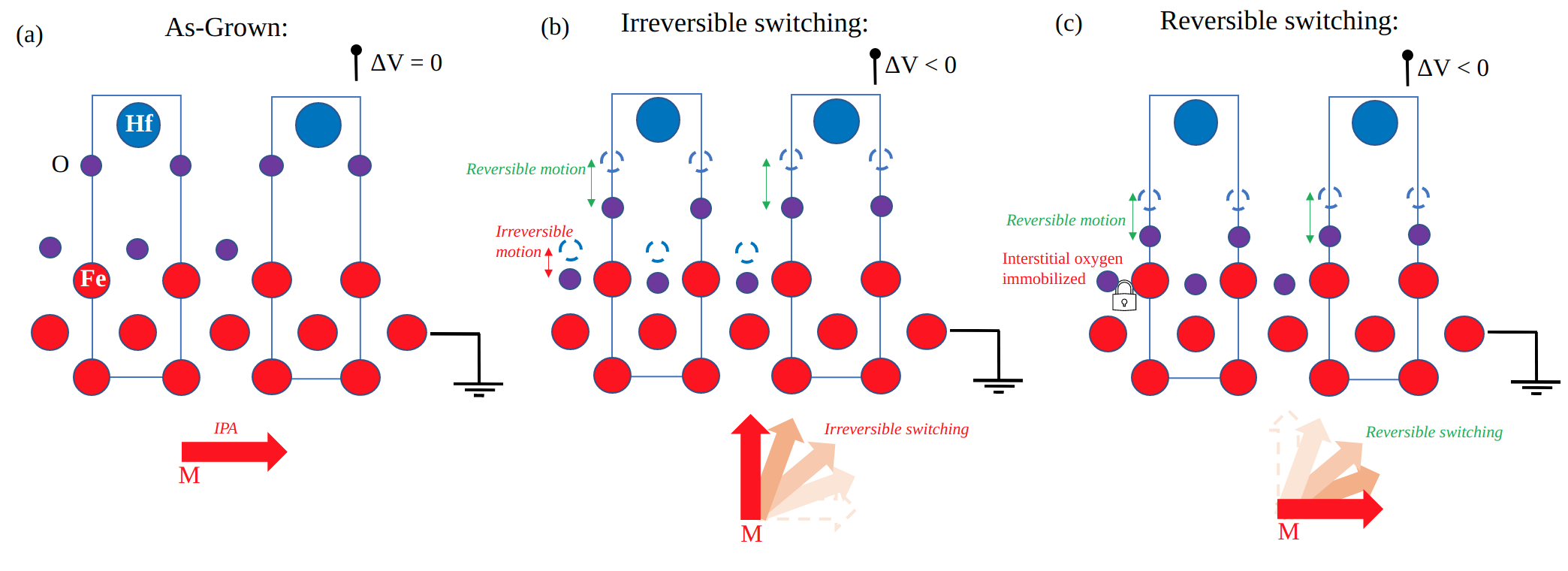}
  \caption{Hypothesis for the mechanism governing the different magneto-ionic regimes in CoFeB/HfO$_2$ multilayers. (a) Shows the ground state of the system with both pure and mixed interfaces. (b) Irreversible magnetization switching via interstitial oxygen shifts. (c) Reversible magnetization switching via frontal oxygen shifts. The bottom cartoon in all three panels represents the magnetization direction and the switching process.}
  \label{micro_mech}
\end{figure*}

Considering the results presented above, we propose the following hypothesis for the appearance of different magneto-ionic regimes in HfO$_2/$CoFeB. As can be seen from Fig.~\ref{fig:HfO_2/Fe_MAE}-(a) and Fig.~\ref{fig:DFT_Cells}, the pure Fe/HfO$_2$ surface with frontally aligned oxygens appears to be both the more stable structure while at the same time displaying IP magnetic anisotropy. This is in accordance with the experimental data, where samples in the as-grown form displayed IP magnetic anisotropy \cite{Pachat}. We also point out how the ground state of the Fe/HfO$_2$ surface with both frontal and interstitial oxygens also displays IP magnetic anisotropy (Fig.~\ref{fig:HfO_2/Fe_Ox_mixed}-(a)), leading us to hypothesize that the actual as-grown surface displays these 2 interfacial oxygen arrangements (see Fig.~\ref{micro_mech}-(a)). As observed in \cite{Pachat}, the application of a gate negative voltage across the Ta/CoFeB/HfO$_2$ favouring the diffusion of oxygen species into the CoFeB layer, results in an irreversible SRT from IPA to PMA. We hypothesize that in this first regime the frontal oxygen atoms, which are easier to mobilize (Fig.~\ref{fig:HfO_2/Fe_mixed_energy_cost}-(a)), are too far away to contribute to the anisotropy manipulation (Fig.~\ref{fig:HfO_2/Fe_Ox_mixed}-(b)). In contrast, the (harder to mobilize) interstitial oxygen atoms of the mixed surface have a decisive impact on the anisotropy of the system and can induce PMA with very small displacements (Fig.~\ref{fig:HfO_2/Fe_Ox_mixed}-(d)). We therefore conclude that the initial irreversible anisotropy change in Ta/CoFeB/HfO$_2$ is due to the irreversibility of the shifts of oxygens located at the interstitial site of mixed interface configurations (Fig.~\ref{micro_mech}-(b)). Once PMA is achieved (Fig.~\ref{fig:HfO_2/Fe_Ox_mixed}-(d)), we hypothesize that the interstitial oxygen atoms become essentially immobile under the effect of the electric field because of the screening effects of the conductive material \cite{Zhang1999}. The experimental data \cite{Pachat} shows that the application of the gate voltage beyond the PMA state has the effect of pushing the magnetic anisotropy easy axis in-plane, albeit in a reversible way. Our results suggest that this switch to a reversible behavior beyond PMA could be explained by the following: the continued application of the gate voltage has the effect of mobilizing the remaining frontal oxygen atoms at the pure and mixed sites, which move closer to the surface and restore in-plane magnetic anisotropy (Fig.~\ref{fig:HfO_2/Fe_Ox_mixed}-(e)). Mobilizing frontal oxygen species is easier (Fig.~\ref{fig:HfO_2/Fe_mixed_energy_cost}) leading us to believe that the reversibility of the regime presented in ref.\cite{Pachat} could be attributed to the reversible shifts of the frontal oxygen species of both the pure and the mixed interface (Fig.~\ref{micro_mech}-(c)). As shown in Fig.~\ref{fig:HfO_2/Fe_Ox_mixed}-(c), the transition PMA $\rightarrow$ IPA can also be achieved by shifting the interstitial oxygens deeper inside the ferromagnetic layer. We do not attribute the manipulation of anisotropy beyond initial PMA to these oxygen species because, once the anisotropy is shifted in-plane by means of the interstitial oxygen species pushed deeper in the sample, it is impossible to manipulate the magnetic anisotropy of the system by shifting the frontal oxygen species closer to the Fe surface (as displayed in Fig.~\ref{fig:HfO_2/Fe_mixed_energy_cost}-(b)).

\section{\label{Conclusion}Conclusion}
In this work we performed an ab-initio analysis of the interplay between oxygen ionic mobility and anisotropy manipulation in $2$ FM/Oxide interfaces in order to formulate a hypothesis for the appearance of magneto-ionic regimes in Ta/CoFeB/HfO$_2$ stacks \cite{Pachat}. We found out that the different nature of the oxide at the interface plays an important role in determining the optimal interfacial oxygen geometry. In particular, we discovered how the pure Fe/HfO$_2$ interface displays a preferential frontal oxygen alignment which corresponds to an IP magnetic anisotropy. We observed how frontal oxygen mobility can induce PMA \cite{Hallal2013}. The inclusion of oxygen species in additional interstitial sites was investigated in order to determine their role in the appearance of magneto-ionic regimes \cite{Pachat}. We have shown how in these so called mixed surfaces, the ionic mobility of frontal oxygen species is energetically more favorable than the mobility of interstitial ones. We have also shown how the interplay of mobility between these 2 different oxygen species can change the magnetic anisotropy of the sample. We conclude that the irreversibility of the transition between IPA (under-oxidised) and PMA (optimally oxidised) is mainly due to the interaction of the interstitial oxygen of the mixed surface and that the reversibility of the second regime is due to the mobility of the frontal oxygen species of the pure and mixed surface (Fig.~\ref{micro_mech}). We point out how the Ta/CoFeB/HfO$_2$ multilayer in \cite{Pachat} is composed by amorphous materials that do not display the ordered structure of our unit cells. In spite of this difference, we find qualitative agreement between our results and the ones reported in \cite{Pachat} and are therefore led to believe that anisotropy manipulation in these kinds of systems may be a consequence of relative FM/oxygen positioning localized at the interface. 
To further probe the validity of the hypothesis, one could analyze the relative range of magneto-ionic regimes in Ta/CoFeB/HfO$_2$ samples with different degrees of atmospheric oxygen interaction. We predict that in the extreme case where there has been little to no exposure to atmospheric oxygen should result in a purely reversible system. 
In addition, a comparison with crystalline and policrystalline structures could prove useful in understanding the role played by the amorphous nature of the materials. Understanding the precise nature of the interplay between ionic mobility and magnetic property tuning could prove very useful for the optimization of highly energy efficient read/write mechanisms for next-generation memory storage devices. 

\section{Acknowledgement}
This project has received funding from the European Union’s Horizon 2020 research and innovation programme under the Marie Skłodowska-Curie grant agreement No. 860060 “Magnetism and the effect of Electric Field” (MagnEFi). Computational resources were provided by HPC@POLITO, a project of
Academic Computing within the Department of Control and Computer
Engineering at the Politecnico di Torino (http://hpc.polito.it)


\begin{thebibliography}{34}%
\makeatletter
\providecommand \@ifxundefined [1]{%
 \@ifx{#1\undefined}
}%
\providecommand \@ifnum [1]{%
 \ifnum #1\expandafter \@firstoftwo
 \else \expandafter \@secondoftwo
 \fi
}%
\providecommand \@ifx [1]{%
 \ifx #1\expandafter \@firstoftwo
 \else \expandafter \@secondoftwo
 \fi
}%
\providecommand \natexlab [1]{#1}%
\providecommand \enquote  [1]{``#1''}%
\providecommand \bibnamefont  [1]{#1}%
\providecommand \bibfnamefont [1]{#1}%
\providecommand \citenamefont [1]{#1}%
\providecommand \href@noop [0]{\@secondoftwo}%
\providecommand \href [0]{\begingroup \@sanitize@url \@href}%
\providecommand \@href[1]{\@@startlink{#1}\@@href}%
\providecommand \@@href[1]{\endgroup#1\@@endlink}%
\providecommand \@sanitize@url [0]{\catcode `\\12\catcode `\$12\catcode
  `\&12\catcode `\#12\catcode `\^12\catcode `\_12\catcode `\%12\relax}%
\providecommand \@@startlink[1]{}%
\providecommand \@@endlink[0]{}%
\providecommand \url  [0]{\begingroup\@sanitize@url \@url }%
\providecommand \@url [1]{\endgroup\@href {#1}{\urlprefix }}%
\providecommand \urlprefix  [0]{URL }%
\providecommand \Eprint [0]{\href }%
\providecommand \doibase [0]{https://doi.org/}%
\providecommand \selectlanguage [0]{\@gobble}%
\providecommand \bibinfo  [0]{\@secondoftwo}%
\providecommand \bibfield  [0]{\@secondoftwo}%
\providecommand \translation [1]{[#1]}%
\providecommand \BibitemOpen [0]{}%
\providecommand \bibitemStop [0]{}%
\providecommand \bibitemNoStop [0]{.\EOS\space}%
\providecommand \EOS [0]{\spacefactor3000\relax}%
\providecommand \BibitemShut  [1]{\csname bibitem#1\endcsname}%
\let\auto@bib@innerbib\@empty
\bibitem [{\citenamefont {One}\ \emph {et~al.}(2021)\citenamefont {One},
  \citenamefont {B{\'{e}}a}, \citenamefont {Mican}, \citenamefont {Joldos},
  \citenamefont {Veiga}, \citenamefont {Dieny}, \citenamefont
  {Buda-Prejbeanu},\ and\ \citenamefont {Tiusan}}]{One2021}%
  \BibitemOpen
  \bibfield  {author} {\bibinfo {author} {\bibfnamefont {R.-A.}\ \bibnamefont
  {One}}, \bibinfo {author} {\bibfnamefont {H.}~\bibnamefont {B{\'{e}}a}},
  \bibinfo {author} {\bibfnamefont {S.}~\bibnamefont {Mican}}, \bibinfo
  {author} {\bibfnamefont {M.}~\bibnamefont {Joldos}}, \bibinfo {author}
  {\bibfnamefont {P.~B.}\ \bibnamefont {Veiga}}, \bibinfo {author}
  {\bibfnamefont {B.}~\bibnamefont {Dieny}}, \bibinfo {author} {\bibfnamefont
  {L.~D.}\ \bibnamefont {Buda-Prejbeanu}},\ and\ \bibinfo {author}
  {\bibfnamefont {C.}~\bibnamefont {Tiusan}},\ }\bibfield  {title} {\bibinfo
  {title} {Route towards efficient magnetization reversal driven by voltage
  control of magnetic anisotropy},\ }\href
  {https://doi.org/10.1038/s41598-021-88408-z} {\bibfield  {journal} {\bibinfo
  {journal} {Sci. Rep.}\ }\textbf {\bibinfo {volume} {11}},\ \bibinfo {pages}
  {8801} (\bibinfo {year} {2021})}\BibitemShut {NoStop}%
\bibitem [{\citenamefont {Suwardy}\ \emph {et~al.}(2019)\citenamefont
  {Suwardy}, \citenamefont {Goto}, \citenamefont {Suzuki},\ and\ \citenamefont
  {Miwa}}]{SUW-19}%
  \BibitemOpen
  \bibfield  {author} {\bibinfo {author} {\bibfnamefont {J.}~\bibnamefont
  {Suwardy}}, \bibinfo {author} {\bibfnamefont {M.}~\bibnamefont {Goto}},
  \bibinfo {author} {\bibfnamefont {Y.}~\bibnamefont {Suzuki}},\ and\ \bibinfo
  {author} {\bibfnamefont {S.}~\bibnamefont {Miwa}},\ }\bibfield  {title}
  {\bibinfo {title} {{Voltage-controlled magnetic anisotropy and
  Dzyaloshinskii-Moriya interactions in {CoNi}/{MgO} and {CoNi}/Pd/{MgO}}},\
  }\href {https://doi.org/10.7567/1347-4065/ab21a6} {\bibfield  {journal}
  {\bibinfo  {journal} {Jpn. J. Appl. Phys.}\ }\textbf {\bibinfo {volume}
  {58}},\ \bibinfo {pages} {60917} (\bibinfo {year} {2019})}\BibitemShut
  {NoStop}%
\bibitem [{\citenamefont {Dieny}\ and\ \citenamefont {Chshiev}(2017)}]{DIE-17}%
  \BibitemOpen
  \bibfield  {author} {\bibinfo {author} {\bibfnamefont {B.}~\bibnamefont
  {Dieny}}\ and\ \bibinfo {author} {\bibfnamefont {M.}~\bibnamefont
  {Chshiev}},\ }\bibfield  {title} {\bibinfo {title} {{Perpendicular magnetic
  anisotropy at transition metal/oxide interfaces and applications}},\ }\href
  {https://doi.org/10.1103/revmodphys.89.025008} {\bibfield  {journal}
  {\bibinfo  {journal} {Rev. Mod. Phys.}\ }\textbf {\bibinfo {volume} {89}},\
  \bibinfo {pages} {025008} (\bibinfo {year} {2017})}\BibitemShut {NoStop}%
\bibitem [{\citenamefont {Hallal}\ \emph {et~al.}(2013)\citenamefont {Hallal},
  \citenamefont {Yang}, \citenamefont {Dieny},\ and\ \citenamefont
  {Chshiev}}]{Hallal2013}%
  \BibitemOpen
  \bibfield  {author} {\bibinfo {author} {\bibfnamefont {A.}~\bibnamefont
  {Hallal}}, \bibinfo {author} {\bibfnamefont {H.~X.}\ \bibnamefont {Yang}},
  \bibinfo {author} {\bibfnamefont {B.}~\bibnamefont {Dieny}},\ and\ \bibinfo
  {author} {\bibfnamefont {M.}~\bibnamefont {Chshiev}},\ }\bibfield  {title}
  {\bibinfo {title} {{Anatomy of perpendicular magnetic anisotropy in Fe/MgO
  magnetic tunnel junctions: First-principles insight}},\ }\href
  {https://doi.org/10.1103/PhysRevB.88.184423} {\bibfield  {journal} {\bibinfo
  {journal} {Phys. Rev. B}\ }\textbf {\bibinfo {volume} {88}},\ \bibinfo
  {pages} {1} (\bibinfo {year} {2013})}\BibitemShut {NoStop}%
\bibitem [{\citenamefont {Belabbes}\ \emph {et~al.}(2016)\citenamefont
  {Belabbes}, \citenamefont {Bihlmayer}, \citenamefont {Bl{\"{u}}gel},\ and\
  \citenamefont {Manchon}}]{Belabbes2016}%
  \BibitemOpen
  \bibfield  {author} {\bibinfo {author} {\bibfnamefont {A.}~\bibnamefont
  {Belabbes}}, \bibinfo {author} {\bibfnamefont {G.}~\bibnamefont {Bihlmayer}},
  \bibinfo {author} {\bibfnamefont {S.}~\bibnamefont {Bl{\"{u}}gel}},\ and\
  \bibinfo {author} {\bibfnamefont {A.}~\bibnamefont {Manchon}},\ }\bibfield
  {title} {\bibinfo {title} {{Oxygen-enabled control of Dzyaloshinskii-Moriya
  Interaction in ultra-thin magnetic films}},\ }\href
  {https://doi.org/10.1038/srep24634} {\bibfield  {journal} {\bibinfo
  {journal} {Sci. Rep.}\ }\textbf {\bibinfo {volume} {6}},\ \bibinfo {pages}
  {24634} (\bibinfo {year} {2016})}\BibitemShut {NoStop}%
\bibitem [{\citenamefont {Ibrahim}\ \emph {et~al.}(2016)\citenamefont
  {Ibrahim}, \citenamefont {Yang}, \citenamefont {Hallal}, \citenamefont
  {Dieny},\ and\ \citenamefont {Chshiev}}]{IBR-16}%
  \BibitemOpen
  \bibfield  {author} {\bibinfo {author} {\bibfnamefont {F.}~\bibnamefont
  {Ibrahim}}, \bibinfo {author} {\bibfnamefont {H.~X.}\ \bibnamefont {Yang}},
  \bibinfo {author} {\bibfnamefont {A.}~\bibnamefont {Hallal}}, \bibinfo
  {author} {\bibfnamefont {B.}~\bibnamefont {Dieny}},\ and\ \bibinfo {author}
  {\bibfnamefont {M.}~\bibnamefont {Chshiev}},\ }\bibfield  {title} {\bibinfo
  {title} {{Anatomy of electric field control of perpendicular magnetic
  anisotropy at Fe/{MgO} interfaces}},\ }\href
  {https://doi.org/10.1103/physrevb.93.014429} {\bibfield  {journal} {\bibinfo
  {journal} {Phys. Rev. B}\ }\textbf {\bibinfo {volume} {93}},\ \bibinfo
  {pages} {184423} (\bibinfo {year} {2016})}\BibitemShut {NoStop}%
\bibitem [{\citenamefont {Huang}\ \emph {et~al.}(2013)\citenamefont {Huang},
  \citenamefont {Stolichnov}, \citenamefont {Bernand-Mantel}, \citenamefont
  {Borrel}, \citenamefont {Auffret}, \citenamefont {Gaudin}, \citenamefont
  {Boulle}, \citenamefont {Pizzini}, \citenamefont {Ranno}, \citenamefont
  {Diez},\ and\ \citenamefont {Setter}}]{HUA-13}%
  \BibitemOpen
  \bibfield  {author} {\bibinfo {author} {\bibfnamefont {Z.}~\bibnamefont
  {Huang}}, \bibinfo {author} {\bibfnamefont {I.}~\bibnamefont {Stolichnov}},
  \bibinfo {author} {\bibfnamefont {A.}~\bibnamefont {Bernand-Mantel}},
  \bibinfo {author} {\bibfnamefont {J.}~\bibnamefont {Borrel}}, \bibinfo
  {author} {\bibfnamefont {S.}~\bibnamefont {Auffret}}, \bibinfo {author}
  {\bibfnamefont {G.}~\bibnamefont {Gaudin}}, \bibinfo {author} {\bibfnamefont
  {O.}~\bibnamefont {Boulle}}, \bibinfo {author} {\bibfnamefont
  {S.}~\bibnamefont {Pizzini}}, \bibinfo {author} {\bibfnamefont
  {L.}~\bibnamefont {Ranno}}, \bibinfo {author} {\bibfnamefont {L.~H.}\
  \bibnamefont {Diez}},\ and\ \bibinfo {author} {\bibfnamefont
  {N.}~\bibnamefont {Setter}},\ }\bibfield  {title} {\bibinfo {title}
  {{Ferroelectric control of magnetic domains in ultra-thin cobalt layers}},\
  }\href {https://doi.org/10.1063/1.4833495} {\bibfield  {journal} {\bibinfo
  {journal} {Appl. Phys. Lett.}\ }\textbf {\bibinfo {volume} {103}},\ \bibinfo
  {pages} {222902} (\bibinfo {year} {2013})}\BibitemShut {NoStop}%
\bibitem [{\citenamefont {Niranjan}\ \emph {et~al.}(2010)\citenamefont
  {Niranjan}, \citenamefont {Duan}, \citenamefont {Jaswal},\ and\ \citenamefont
  {Tsymbal}}]{Niranjan2010}%
  \BibitemOpen
  \bibfield  {author} {\bibinfo {author} {\bibfnamefont {M.~K.}\ \bibnamefont
  {Niranjan}}, \bibinfo {author} {\bibfnamefont {C.~G.}\ \bibnamefont {Duan}},
  \bibinfo {author} {\bibfnamefont {S.~S.}\ \bibnamefont {Jaswal}},\ and\
  \bibinfo {author} {\bibfnamefont {E.~Y.}\ \bibnamefont {Tsymbal}},\
  }\bibfield  {title} {\bibinfo {title} {{Electric field effect on
  magnetization at the Fe/MgO(001) interface}},\ }\href
  {https://doi.org/10.1063/1.3443658} {\bibfield  {journal} {\bibinfo
  {journal} {Appl. Phys. Lett.}\ }\textbf {\bibinfo {volume} {96}},\ \bibinfo
  {pages} {222504} (\bibinfo {year} {2010})}\BibitemShut {NoStop}%
\bibitem [{\citenamefont {Monso}\ \emph {et~al.}(2002)\citenamefont {Monso},
  \citenamefont {Rodmacq}, \citenamefont {Auffret}, \citenamefont {Casali},
  \citenamefont {Fettar}, \citenamefont {Gilles}, \citenamefont {Dieny},\ and\
  \citenamefont {Boyer}}]{Monso2002}%
  \BibitemOpen
  \bibfield  {author} {\bibinfo {author} {\bibfnamefont {S.}~\bibnamefont
  {Monso}}, \bibinfo {author} {\bibfnamefont {B.}~\bibnamefont {Rodmacq}},
  \bibinfo {author} {\bibfnamefont {S.}~\bibnamefont {Auffret}}, \bibinfo
  {author} {\bibfnamefont {G.}~\bibnamefont {Casali}}, \bibinfo {author}
  {\bibfnamefont {F.}~\bibnamefont {Fettar}}, \bibinfo {author} {\bibfnamefont
  {B.}~\bibnamefont {Gilles}}, \bibinfo {author} {\bibfnamefont
  {B.}~\bibnamefont {Dieny}},\ and\ \bibinfo {author} {\bibfnamefont
  {P.}~\bibnamefont {Boyer}},\ }\bibfield  {title} {\bibinfo {title}
  {{Crossover from in-plane to perpendicular anisotropy in Pt/CoFe/AlO x
  sandwiches as a function of Al oxidation: A very accurate control of the
  oxidation of tunnel barriers}},\ }\href {https://doi.org/10.1063/1.1483122}
  {\bibfield  {journal} {\bibinfo  {journal} {Appl. Phys. Lett.}\ }\textbf
  {\bibinfo {volume} {80}},\ \bibinfo {pages} {4157} (\bibinfo {year}
  {2002})}\BibitemShut {NoStop}%
\bibitem [{\citenamefont {Bauer}\ \emph {et~al.}(2015)\citenamefont {Bauer},
  \citenamefont {Yao}, \citenamefont {Tan}, \citenamefont {Agrawal},
  \citenamefont {Emori}, \citenamefont {Tuller}, \citenamefont {{Van Dijken}},\
  and\ \citenamefont {Beach}}]{Bauer2015}%
  \BibitemOpen
  \bibfield  {author} {\bibinfo {author} {\bibfnamefont {U.}~\bibnamefont
  {Bauer}}, \bibinfo {author} {\bibfnamefont {L.}~\bibnamefont {Yao}}, \bibinfo
  {author} {\bibfnamefont {A.~J.}\ \bibnamefont {Tan}}, \bibinfo {author}
  {\bibfnamefont {P.}~\bibnamefont {Agrawal}}, \bibinfo {author} {\bibfnamefont
  {S.}~\bibnamefont {Emori}}, \bibinfo {author} {\bibfnamefont {H.~L.}\
  \bibnamefont {Tuller}}, \bibinfo {author} {\bibfnamefont {S.}~\bibnamefont
  {{Van Dijken}}},\ and\ \bibinfo {author} {\bibfnamefont {G.~S.}\ \bibnamefont
  {Beach}},\ }\bibfield  {title} {\bibinfo {title} {{Magneto-ionic control of
  interfacial magnetism}},\ }\href {https://doi.org/10.1038/nmat4134}
  {\bibfield  {journal} {\bibinfo  {journal} {Nat. Mater.}\ }\textbf {\bibinfo
  {volume} {14}},\ \bibinfo {pages} {174} (\bibinfo {year} {2015})}\BibitemShut
  {NoStop}%
\bibitem [{\citenamefont {Fassatoui}\ \emph {et~al.}(2020)\citenamefont
  {Fassatoui}, \citenamefont {Garcia}, \citenamefont {Ranno}, \citenamefont
  {Vogel}, \citenamefont {Bernand-Mantel}, \citenamefont {B{\'{e}}a},
  \citenamefont {Pizzini},\ and\ \citenamefont {Pizzini}}]{Fassatoui2020}%
  \BibitemOpen
  \bibfield  {author} {\bibinfo {author} {\bibfnamefont {A.}~\bibnamefont
  {Fassatoui}}, \bibinfo {author} {\bibfnamefont {J.~P.}\ \bibnamefont
  {Garcia}}, \bibinfo {author} {\bibfnamefont {L.}~\bibnamefont {Ranno}},
  \bibinfo {author} {\bibfnamefont {J.}~\bibnamefont {Vogel}}, \bibinfo
  {author} {\bibfnamefont {A.}~\bibnamefont {Bernand-Mantel}}, \bibinfo
  {author} {\bibfnamefont {H.}~\bibnamefont {B{\'{e}}a}}, \bibinfo {author}
  {\bibfnamefont {S.}~\bibnamefont {Pizzini}},\ and\ \bibinfo {author}
  {\bibfnamefont {S.}~\bibnamefont {Pizzini}},\ }\bibfield  {title} {\bibinfo
  {title} {{Reversible and Irreversible Voltage Manipulation of Interfacial
  Magnetic Anisotropy in Pt / Co /Oxide Multilayers}},\ }\href
  {https://doi.org/10.1103/PhysRevApplied.14.064041} {\bibfield  {journal}
  {\bibinfo  {journal} {Phys. Rev. Applied}\ }\textbf {\bibinfo {volume}
  {14}},\ \bibinfo {pages} {064041} (\bibinfo {year} {2020})}\BibitemShut
  {NoStop}%
\bibitem [{\citenamefont {Flynn}(1962)}]{Flynn1962}%
  \BibitemOpen
  \bibfield  {author} {\bibinfo {author} {\bibfnamefont {H.~G.}\ \bibnamefont
  {Flynn}},\ }\bibfield  {title} {\bibinfo {title} {{Transport Phenomena}},\
  }\href {https://doi.org/10.1063/1.3058263} {\bibfield  {journal} {\bibinfo
  {journal} {Phys. Today}\ }\textbf {\bibinfo {volume} {15}},\ \bibinfo {pages}
  {96} (\bibinfo {year} {1962})}\BibitemShut {NoStop}%
\bibitem [{\citenamefont {Wu}\ \emph {et~al.}(2017)\citenamefont {Wu},
  \citenamefont {Wang}, \citenamefont {Wei},\ and\ \citenamefont
  {Li}}]{Wu2017}%
  \BibitemOpen
  \bibfield  {author} {\bibinfo {author} {\bibfnamefont {N.}~\bibnamefont
  {Wu}}, \bibinfo {author} {\bibfnamefont {W.}~\bibnamefont {Wang}}, \bibinfo
  {author} {\bibfnamefont {Y.}~\bibnamefont {Wei}},\ and\ \bibinfo {author}
  {\bibfnamefont {T.}~\bibnamefont {Li}},\ }\bibfield  {title} {\bibinfo
  {title} {{Studies on the effect of nano-sized MgO in magnesium-ion conducting
  gel polymer electrolyte for rechargeable magnesium batteries}},\ }\href
  {https://doi.org/10.3390/en10081215} {\bibfield  {journal} {\bibinfo
  {journal} {Energies}\ }\textbf {\bibinfo {volume} {10}},\ \bibinfo {pages}
  {1215} (\bibinfo {year} {2017})}\BibitemShut {NoStop}%
\bibitem [{\citenamefont {Liu}\ \emph {et~al.}(2014)\citenamefont {Liu},
  \citenamefont {Zhang}, \citenamefont {Cai},\ and\ \citenamefont
  {Pan}}]{Liu2014}%
  \BibitemOpen
  \bibfield  {author} {\bibinfo {author} {\bibfnamefont {T.}~\bibnamefont
  {Liu}}, \bibinfo {author} {\bibfnamefont {Y.}~\bibnamefont {Zhang}}, \bibinfo
  {author} {\bibfnamefont {J.~W.}\ \bibnamefont {Cai}},\ and\ \bibinfo {author}
  {\bibfnamefont {H.~Y.}\ \bibnamefont {Pan}},\ }\bibfield  {title} {\bibinfo
  {title} {{Thermally robust Mo/CoFeB/MgO trilayers with strong perpendicular
  magnetic anisotropy}},\ }\href {https://doi.org/10.1038/srep05895} {\bibfield
   {journal} {\bibinfo  {journal} {Sci. Rep.}\ }\textbf {\bibinfo {volume}
  {4}},\ \bibinfo {pages} {5895} (\bibinfo {year} {2014})}\BibitemShut
  {NoStop}%
\bibitem [{\citenamefont {Ikeda}\ \emph {et~al.}(2010)\citenamefont {Ikeda},
  \citenamefont {Miura}, \citenamefont {Yamamoto}, \citenamefont {Mizunuma},
  \citenamefont {Gan}, \citenamefont {Endo}, \citenamefont {Kanai},
  \citenamefont {Hayakawa}, \citenamefont {Matsukura},\ and\ \citenamefont
  {Ohno}}]{Ikeda2010}%
  \BibitemOpen
  \bibfield  {author} {\bibinfo {author} {\bibfnamefont {S.}~\bibnamefont
  {Ikeda}}, \bibinfo {author} {\bibfnamefont {K.}~\bibnamefont {Miura}},
  \bibinfo {author} {\bibfnamefont {H.}~\bibnamefont {Yamamoto}}, \bibinfo
  {author} {\bibfnamefont {K.}~\bibnamefont {Mizunuma}}, \bibinfo {author}
  {\bibfnamefont {H.~D.}\ \bibnamefont {Gan}}, \bibinfo {author} {\bibfnamefont
  {M.}~\bibnamefont {Endo}}, \bibinfo {author} {\bibfnamefont {S.}~\bibnamefont
  {Kanai}}, \bibinfo {author} {\bibfnamefont {J.}~\bibnamefont {Hayakawa}},
  \bibinfo {author} {\bibfnamefont {F.}~\bibnamefont {Matsukura}},\ and\
  \bibinfo {author} {\bibfnamefont {H.}~\bibnamefont {Ohno}},\ }\bibfield
  {title} {\bibinfo {title} {{A perpendicular-anisotropy CoFeB-MgO magnetic
  tunnel junction}},\ }\href {https://doi.org/10.1038/nmat2804} {\bibfield
  {journal} {\bibinfo  {journal} {Nat. Mater.}\ }\textbf {\bibinfo {volume}
  {9}},\ \bibinfo {pages} {721} (\bibinfo {year} {2010})}\BibitemShut {NoStop}%
\bibitem [{\citenamefont {Wang}\ \emph {et~al.}(2020)\citenamefont {Wang},
  \citenamefont {Li}, \citenamefont {Sasaki}, \citenamefont {Wong},
  \citenamefont {Yu}, \citenamefont {Peng}, \citenamefont {Zhao}, \citenamefont
  {Ohkubo}, \citenamefont {Hono}, \citenamefont {Zhao},\ and\ \citenamefont
  {Wang}}]{Wang2020}%
  \BibitemOpen
  \bibfield  {author} {\bibinfo {author} {\bibfnamefont {L.~Z.}\ \bibnamefont
  {Wang}}, \bibinfo {author} {\bibfnamefont {X.}~\bibnamefont {Li}}, \bibinfo
  {author} {\bibfnamefont {T.}~\bibnamefont {Sasaki}}, \bibinfo {author}
  {\bibfnamefont {K.}~\bibnamefont {Wong}}, \bibinfo {author} {\bibfnamefont
  {G.~Q.}\ \bibnamefont {Yu}}, \bibinfo {author} {\bibfnamefont {S.~Z.}\
  \bibnamefont {Peng}}, \bibinfo {author} {\bibfnamefont {C.}~\bibnamefont
  {Zhao}}, \bibinfo {author} {\bibfnamefont {T.}~\bibnamefont {Ohkubo}},
  \bibinfo {author} {\bibfnamefont {K.}~\bibnamefont {Hono}}, \bibinfo {author}
  {\bibfnamefont {W.~S.}\ \bibnamefont {Zhao}},\ and\ \bibinfo {author}
  {\bibfnamefont {K.~L.}\ \bibnamefont {Wang}},\ }\bibfield  {title} {\bibinfo
  {title} {{High voltage-controlled magnetic anisotropy and interface
  magnetoelectric effect in sputtered multilayers annealed at high
  temperatures}},\ }\href {https://doi.org/10.1007/s11433-019-1524-y}
  {\bibfield  {journal} {\bibinfo  {journal} {Science China: Physics, Mechanics
  and Astronomy}\ }\textbf {\bibinfo {volume} {63}},\ \bibinfo {pages} {277512}
  (\bibinfo {year} {2020})}\BibitemShut {NoStop}%
\bibitem [{\citenamefont {Pachat}\ \emph {et~al.}(2021)\citenamefont {Pachat},
  \citenamefont {Ourdani}, \citenamefont {van~der Jagt}, \citenamefont
  {Syskaki}, \citenamefont {Pietro}, \citenamefont {Roussign{\'{e}}},
  \citenamefont {Ono}, \citenamefont {Gabor}, \citenamefont {Ch{\'{e}}rif},
  \citenamefont {Durin}, \citenamefont {Langer}, \citenamefont {Belmeguenai},
  \citenamefont {Ravelosona},\ and\ \citenamefont {Diez}}]{Pachat}%
  \BibitemOpen
  \bibfield  {author} {\bibinfo {author} {\bibfnamefont {R.}~\bibnamefont
  {Pachat}}, \bibinfo {author} {\bibfnamefont {D.}~\bibnamefont {Ourdani}},
  \bibinfo {author} {\bibfnamefont {J.}~\bibnamefont {van~der Jagt}}, \bibinfo
  {author} {\bibfnamefont {M.-A.}\ \bibnamefont {Syskaki}}, \bibinfo {author}
  {\bibfnamefont {A.~D.}\ \bibnamefont {Pietro}}, \bibinfo {author}
  {\bibfnamefont {Y.}~\bibnamefont {Roussign{\'{e}}}}, \bibinfo {author}
  {\bibfnamefont {S.}~\bibnamefont {Ono}}, \bibinfo {author} {\bibfnamefont
  {M.}~\bibnamefont {Gabor}}, \bibinfo {author} {\bibfnamefont
  {M.}~\bibnamefont {Ch{\'{e}}rif}}, \bibinfo {author} {\bibfnamefont
  {G.}~\bibnamefont {Durin}}, \bibinfo {author} {\bibfnamefont
  {J.}~\bibnamefont {Langer}}, \bibinfo {author} {\bibfnamefont
  {M.}~\bibnamefont {Belmeguenai}}, \bibinfo {author} {\bibfnamefont
  {D.}~\bibnamefont {Ravelosona}},\ and\ \bibinfo {author} {\bibfnamefont
  {L.~H.}\ \bibnamefont {Diez}},\ }\bibfield  {title} {\bibinfo {title}
  {Multiple magnetoionic regimes in ta/co20fe60b20/{HfO}2},\ }\href
  {https://doi.org/10.1103/physrevapplied.15.064055} {\bibfield  {journal}
  {\bibinfo  {journal} {Phys. Rev. Applied}\ }\textbf {\bibinfo {volume}
  {15}},\ \bibinfo {pages} {064055} (\bibinfo {year} {2021})}\BibitemShut
  {NoStop}%
\bibitem [{\citenamefont {Schie}\ \emph {et~al.}(2017)\citenamefont {Schie},
  \citenamefont {M{\"{u}}ller}, \citenamefont {Salinga}, \citenamefont
  {Waser},\ and\ \citenamefont {{De Souza}}}]{Schie2017a}%
  \BibitemOpen
  \bibfield  {author} {\bibinfo {author} {\bibfnamefont {M.}~\bibnamefont
  {Schie}}, \bibinfo {author} {\bibfnamefont {M.~P.}\ \bibnamefont
  {M{\"{u}}ller}}, \bibinfo {author} {\bibfnamefont {M.}~\bibnamefont
  {Salinga}}, \bibinfo {author} {\bibfnamefont {R.}~\bibnamefont {Waser}},\
  and\ \bibinfo {author} {\bibfnamefont {R.~A.}\ \bibnamefont {{De Souza}}},\
  }\bibfield  {title} {\bibinfo {title} {{Ion migration in crystalline and
  amorphous HfOX}},\ }\href {https://doi.org/10.1063/1.4977453} {\bibfield
  {journal} {\bibinfo  {journal} {J. Chem. Phys.}\ }\textbf {\bibinfo {volume}
  {146}},\ \bibinfo {pages} {094508} (\bibinfo {year} {2017})}\BibitemShut
  {NoStop}%
\bibitem [{\citenamefont {Liang}\ \emph {et~al.}(2014)\citenamefont {Liang},
  \citenamefont {Zhang}, \citenamefont {Barate}, \citenamefont {Frougier},
  \citenamefont {Vidal}, \citenamefont {Renucci}, \citenamefont {Xu},
  \citenamefont {Jaffr{\`{e}}s}, \citenamefont {George}, \citenamefont
  {Devaux}, \citenamefont {Hehn}, \citenamefont {Marie}, \citenamefont
  {Mangin}, \citenamefont {Yang}, \citenamefont {Hallal}, \citenamefont
  {Chshiev}, \citenamefont {Amand}, \citenamefont {Liu}, \citenamefont {Liu},
  \citenamefont {Han}, \citenamefont {Wang},\ and\ \citenamefont
  {Lu}}]{Liang2014}%
  \BibitemOpen
  \bibfield  {author} {\bibinfo {author} {\bibfnamefont {S.~H.}\ \bibnamefont
  {Liang}}, \bibinfo {author} {\bibfnamefont {T.~T.}\ \bibnamefont {Zhang}},
  \bibinfo {author} {\bibfnamefont {P.}~\bibnamefont {Barate}}, \bibinfo
  {author} {\bibfnamefont {J.}~\bibnamefont {Frougier}}, \bibinfo {author}
  {\bibfnamefont {M.}~\bibnamefont {Vidal}}, \bibinfo {author} {\bibfnamefont
  {P.}~\bibnamefont {Renucci}}, \bibinfo {author} {\bibfnamefont
  {B.}~\bibnamefont {Xu}}, \bibinfo {author} {\bibfnamefont {H.}~\bibnamefont
  {Jaffr{\`{e}}s}}, \bibinfo {author} {\bibfnamefont {J.~M.}\ \bibnamefont
  {George}}, \bibinfo {author} {\bibfnamefont {X.}~\bibnamefont {Devaux}},
  \bibinfo {author} {\bibfnamefont {M.}~\bibnamefont {Hehn}}, \bibinfo {author}
  {\bibfnamefont {X.}~\bibnamefont {Marie}}, \bibinfo {author} {\bibfnamefont
  {S.}~\bibnamefont {Mangin}}, \bibinfo {author} {\bibfnamefont {H.~X.}\
  \bibnamefont {Yang}}, \bibinfo {author} {\bibfnamefont {A.}~\bibnamefont
  {Hallal}}, \bibinfo {author} {\bibfnamefont {M.}~\bibnamefont {Chshiev}},
  \bibinfo {author} {\bibfnamefont {T.}~\bibnamefont {Amand}}, \bibinfo
  {author} {\bibfnamefont {H.~F.}\ \bibnamefont {Liu}}, \bibinfo {author}
  {\bibfnamefont {D.~P.}\ \bibnamefont {Liu}}, \bibinfo {author} {\bibfnamefont
  {X.~F.}\ \bibnamefont {Han}}, \bibinfo {author} {\bibfnamefont {Z.~G.}\
  \bibnamefont {Wang}},\ and\ \bibinfo {author} {\bibfnamefont
  {Y.}~\bibnamefont {Lu}},\ }\bibfield  {title} {\bibinfo {title} {{Large and
  robust electrical spin injection into GaAs at zero magnetic field using an
  ultrathin CoFeB/MgO injector}},\ }\href
  {https://doi.org/10.1103/PhysRevB.90.085310} {\bibfield  {journal} {\bibinfo
  {journal} {Phys. Rev. B}\ }\textbf {\bibinfo {volume} {90}},\ \bibinfo
  {pages} {085310} (\bibinfo {year} {2014})}\BibitemShut {NoStop}%
\bibitem [{\citenamefont {Andersen}(1975)}]{AND-75}%
  \BibitemOpen
  \bibfield  {author} {\bibinfo {author} {\bibfnamefont {O.~K.}\ \bibnamefont
  {Andersen}},\ }\bibfield  {title} {\bibinfo {title} {Linear methods in band
  theory},\ }\href {https://doi.org/10.1103/physrevb.12.3060} {\bibfield
  {journal} {\bibinfo  {journal} {Phys. Rev. B}\ }\textbf {\bibinfo {volume}
  {12}},\ \bibinfo {pages} {3060} (\bibinfo {year} {1975})}\BibitemShut
  {NoStop}%
\bibitem [{FLE()}]{FLEUR}%
  \BibitemOpen
  \href {www.flapw.de} {\bibinfo {title} {{Fleur code :
  www.flapw.de}}}\BibitemShut {NoStop}%
\bibitem [{\citenamefont {{Perdew}}\ \emph {et~al.}(1996)\citenamefont
  {{Perdew}}, \citenamefont {{Ernzerhof}},\ and\ \citenamefont
  {{Burke}}}]{1996JChPh.105.9982P}%
  \BibitemOpen
  \bibfield  {author} {\bibinfo {author} {\bibfnamefont {J.~P.}\ \bibnamefont
  {{Perdew}}}, \bibinfo {author} {\bibfnamefont {M.}~\bibnamefont
  {{Ernzerhof}}},\ and\ \bibinfo {author} {\bibfnamefont {K.}~\bibnamefont
  {{Burke}}},\ }\bibfield  {title} {\bibinfo {title} {{Rationale for mixing
  exact exchange with density functional approximations}},\ }\href
  {https://doi.org/10.1063/1.472933} {\bibfield  {journal} {\bibinfo  {journal}
  {\jcp}\ }\textbf {\bibinfo {volume} {105}},\ \bibinfo {pages} {9982}
  (\bibinfo {year} {1996})}\BibitemShut {NoStop}%
\bibitem [{\citenamefont {Lin}\ \emph {et~al.}(2020)\citenamefont {Lin},
  \citenamefont {Yang}, \citenamefont {Chen}, \citenamefont {Wu}, \citenamefont
  {Guo}, \citenamefont {Chen}, \citenamefont {Liu}, \citenamefont {Xie},
  \citenamefont {Shu}, \citenamefont {Hui}, \citenamefont {Chow}, \citenamefont
  {Feng}, \citenamefont {Carlotti}, \citenamefont {Tacchi}, \citenamefont
  {Yang},\ and\ \citenamefont {Chen}}]{LIN-20}%
  \BibitemOpen
  \bibfield  {author} {\bibinfo {author} {\bibfnamefont {W.}~\bibnamefont
  {Lin}}, \bibinfo {author} {\bibfnamefont {B.}~\bibnamefont {Yang}}, \bibinfo
  {author} {\bibfnamefont {A.~P.}\ \bibnamefont {Chen}}, \bibinfo {author}
  {\bibfnamefont {X.}~\bibnamefont {Wu}}, \bibinfo {author} {\bibfnamefont
  {R.}~\bibnamefont {Guo}}, \bibinfo {author} {\bibfnamefont {S.}~\bibnamefont
  {Chen}}, \bibinfo {author} {\bibfnamefont {L.}~\bibnamefont {Liu}}, \bibinfo
  {author} {\bibfnamefont {Q.}~\bibnamefont {Xie}}, \bibinfo {author}
  {\bibfnamefont {X.}~\bibnamefont {Shu}}, \bibinfo {author} {\bibfnamefont
  {Y.}~\bibnamefont {Hui}}, \bibinfo {author} {\bibfnamefont {G.~M.}\
  \bibnamefont {Chow}}, \bibinfo {author} {\bibfnamefont {Y.}~\bibnamefont
  {Feng}}, \bibinfo {author} {\bibfnamefont {G.}~\bibnamefont {Carlotti}},
  \bibinfo {author} {\bibfnamefont {S.}~\bibnamefont {Tacchi}}, \bibinfo
  {author} {\bibfnamefont {H.}~\bibnamefont {Yang}},\ and\ \bibinfo {author}
  {\bibfnamefont {J.}~\bibnamefont {Chen}},\ }\bibfield  {title} {\bibinfo
  {title} {{Perpendicular Magnetic Anisotropy and Dzyaloshinskii-Moriya
  Interaction at an Oxide/Ferromagnetic Metal Interface}},\ }\href
  {https://doi.org/10.1103/PhysRevLett.124.217202} {\bibfield  {journal}
  {\bibinfo  {journal} {Phys. Rev. Lett.}\ }\textbf {\bibinfo {volume} {124}},\
  \bibinfo {pages} {217202} (\bibinfo {year} {2020})}\BibitemShut {NoStop}%
\bibitem [{\citenamefont {Yang}\ \emph {et~al.}(2011)\citenamefont {Yang},
  \citenamefont {Chshiev}, \citenamefont {Dieny}, \citenamefont {Lee},
  \citenamefont {Manchon},\ and\ \citenamefont {Shin}}]{YAN-11}%
  \BibitemOpen
  \bibfield  {author} {\bibinfo {author} {\bibfnamefont {H.~X.}\ \bibnamefont
  {Yang}}, \bibinfo {author} {\bibfnamefont {M.}~\bibnamefont {Chshiev}},
  \bibinfo {author} {\bibfnamefont {B.}~\bibnamefont {Dieny}}, \bibinfo
  {author} {\bibfnamefont {J.~H.}\ \bibnamefont {Lee}}, \bibinfo {author}
  {\bibfnamefont {A.}~\bibnamefont {Manchon}},\ and\ \bibinfo {author}
  {\bibfnamefont {K.~H.}\ \bibnamefont {Shin}},\ }\bibfield  {title} {\bibinfo
  {title} {{First-principles investigation of the very large perpendicular
  magnetic anisotropy at Fe$|$MgO and Co$|$MgO interfaces}},\ }\href
  {https://doi.org/10.1103/physrevb.84.054401} {\bibfield  {journal} {\bibinfo
  {journal} {Phys. Rev. B}\ }\textbf {\bibinfo {volume} {84}},\ \bibinfo
  {pages} {054401} (\bibinfo {year} {2011})}\BibitemShut {NoStop}%
\bibitem [{\citenamefont {Strange}\ \emph {et~al.}(1991)\citenamefont
  {Strange}, \citenamefont {Staunton}, \citenamefont {Gy{\"{o}}rffy},\ and\
  \citenamefont {Ebert}}]{Strange1991}%
  \BibitemOpen
  \bibfield  {author} {\bibinfo {author} {\bibfnamefont {P.}~\bibnamefont
  {Strange}}, \bibinfo {author} {\bibfnamefont {J.~B.}\ \bibnamefont
  {Staunton}}, \bibinfo {author} {\bibfnamefont {B.~L.}\ \bibnamefont
  {Gy{\"{o}}rffy}},\ and\ \bibinfo {author} {\bibfnamefont {H.}~\bibnamefont
  {Ebert}},\ }\bibfield  {title} {\bibinfo {title} {{First principles theory of
  magnetocrystalline anisotropy}},\ }\href
  {https://doi.org/10.1016/0921-4526(91)90416-C} {\bibfield  {journal}
  {\bibinfo  {journal} {J. Phys.: Condens. Matter}\ }\textbf {\bibinfo {volume}
  {1}},\ \bibinfo {pages} {3947} (\bibinfo {year} {1991})}\BibitemShut
  {NoStop}%
\bibitem [{\citenamefont {Zimmermann}\ \emph {et~al.}(2019)\citenamefont
  {Zimmermann}, \citenamefont {Bihlmayer}, \citenamefont {B{\"{o}}ttcher},
  \citenamefont {Bouhassoune}, \citenamefont {Lounis}, \citenamefont {Sinova},
  \citenamefont {Heinze}, \citenamefont {Bl{\"{u}}gel},\ and\ \citenamefont
  {Dup{\'{e}}}}]{Zimmermann2019}%
  \BibitemOpen
  \bibfield  {author} {\bibinfo {author} {\bibfnamefont {B.}~\bibnamefont
  {Zimmermann}}, \bibinfo {author} {\bibfnamefont {G.}~\bibnamefont
  {Bihlmayer}}, \bibinfo {author} {\bibfnamefont {M.}~\bibnamefont
  {B{\"{o}}ttcher}}, \bibinfo {author} {\bibfnamefont {M.}~\bibnamefont
  {Bouhassoune}}, \bibinfo {author} {\bibfnamefont {S.}~\bibnamefont {Lounis}},
  \bibinfo {author} {\bibfnamefont {J.}~\bibnamefont {Sinova}}, \bibinfo
  {author} {\bibfnamefont {S.}~\bibnamefont {Heinze}}, \bibinfo {author}
  {\bibfnamefont {S.}~\bibnamefont {Bl{\"{u}}gel}},\ and\ \bibinfo {author}
  {\bibfnamefont {B.}~\bibnamefont {Dup{\'{e}}}},\ }\bibfield  {title}
  {\bibinfo {title} {{Comparison of first-principles methods to extract
  magnetic parameters in ultrathin films: Co/Pt(111)}},\ }\href
  {https://doi.org/10.1103/PhysRevB.99.214426} {\bibfield  {journal} {\bibinfo
  {journal} {Phys. Rev. B}\ }\textbf {\bibinfo {volume} {99}},\ \bibinfo
  {pages} {214426} (\bibinfo {year} {2019})}\BibitemShut {NoStop}%
\bibitem [{\citenamefont {Weinert}\ \emph {et~al.}(1985)\citenamefont
  {Weinert}, \citenamefont {Watson},\ and\ \citenamefont
  {Davenport}}]{Weinert1985}%
  \BibitemOpen
  \bibfield  {author} {\bibinfo {author} {\bibfnamefont {M.}~\bibnamefont
  {Weinert}}, \bibinfo {author} {\bibfnamefont {R.~E.}\ \bibnamefont
  {Watson}},\ and\ \bibinfo {author} {\bibfnamefont {J.~W.}\ \bibnamefont
  {Davenport}},\ }\bibfield  {title} {\bibinfo {title} {{Total-energy
  differences and eigenvalue sums}},\ }\href
  {https://doi.org/10.1103/PhysRevB.32.2115} {\bibfield  {journal} {\bibinfo
  {journal} {Phys. Rev. B}\ }\textbf {\bibinfo {volume} {32}},\ \bibinfo
  {pages} {2115} (\bibinfo {year} {1985})}\BibitemShut {NoStop}%
\bibitem [{\citenamefont {Wang}\ \emph {et~al.}(1996)\citenamefont {Wang},
  \citenamefont {Wang}, \citenamefont {Ruqian},\ and\ \citenamefont
  {Freeman}}]{Wang1996}%
  \BibitemOpen
  \bibfield  {author} {\bibinfo {author} {\bibfnamefont {X.}~\bibnamefont
  {Wang}}, \bibinfo {author} {\bibfnamefont {D.~S.}\ \bibnamefont {Wang}},
  \bibinfo {author} {\bibfnamefont {W.}~\bibnamefont {Ruqian}},\ and\ \bibinfo
  {author} {\bibfnamefont {A.~J.}\ \bibnamefont {Freeman}},\ }\bibfield
  {title} {\bibinfo {title} {{Validity of the force theorem for
  magnetocrystalline anisotropy}},\ }\href
  {https://doi.org/10.1016/0304-8853(95)00936-1} {\bibfield  {journal}
  {\bibinfo  {journal} {J. Magn. Magn. Mater.}\ }\textbf {\bibinfo {volume}
  {159}},\ \bibinfo {pages} {337} (\bibinfo {year} {1996})}\BibitemShut
  {NoStop}%
\bibitem [{\citenamefont {B{\l}o{\'{n}}ski}\ and\ \citenamefont
  {Hafner}(2009)}]{Bonski2009}%
  \BibitemOpen
  \bibfield  {author} {\bibinfo {author} {\bibfnamefont {P.}~\bibnamefont
  {B{\l}o{\'{n}}ski}}\ and\ \bibinfo {author} {\bibfnamefont {J.}~\bibnamefont
  {Hafner}},\ }\bibfield  {title} {\bibinfo {title} {{Density-functional theory
  of the magnetic anisotropy of nanostructures: An assessment of different
  approximations}},\ }\href {https://doi.org/10.1088/0953-8984/21/42/426001}
  {\bibfield  {journal} {\bibinfo  {journal} {J. Phys.: Condens. Matter}\
  }\textbf {\bibinfo {volume} {21}},\ \bibinfo {pages} {426001} (\bibinfo
  {year} {2009})}\BibitemShut {NoStop}%
\bibitem [{\citenamefont {Bruno}(1989)}]{Bruno1989}%
  \BibitemOpen
  \bibfield  {author} {\bibinfo {author} {\bibfnamefont {P.}~\bibnamefont
  {Bruno}},\ }\bibfield  {title} {\bibinfo {title} {{Tight-binding approach to
  the orbital magnetic moment and magnetocrystalline anisotropy of
  transition-metal monolayers}},\ }\href
  {https://doi.org/10.1103/PhysRevB.39.865} {\bibfield  {journal} {\bibinfo
  {journal} {Phys. Rev. B}\ }\textbf {\bibinfo {volume} {39}},\ \bibinfo
  {pages} {865} (\bibinfo {year} {1989})}\BibitemShut {NoStop}%
\bibitem [{\citenamefont {{Van Der Laan}}(1998)}]{VanDerLaan1998}%
  \BibitemOpen
  \bibfield  {author} {\bibinfo {author} {\bibfnamefont {G.}~\bibnamefont {{Van
  Der Laan}}},\ }\bibfield  {title} {\bibinfo {title} {{Microscopic origin of
  magnetocrystalline anisotropy in transition metal thin films}},\ }\href
  {https://doi.org/10.1088/0953-8984/10/14/012} {\bibfield  {journal} {\bibinfo
   {journal} {J. Phys.: Condens. Matter}\ }\textbf {\bibinfo {volume} {10}},\
  \bibinfo {pages} {3239} (\bibinfo {year} {1998})}\BibitemShut {NoStop}%
\bibitem [{\citenamefont {Butler}(2008)}]{Butler2008}%
  \BibitemOpen
  \bibfield  {author} {\bibinfo {author} {\bibfnamefont {W.~H.}\ \bibnamefont
  {Butler}},\ }\bibfield  {title} {\bibinfo {title} {{Tunneling
  magnetoresistance from a symmetry filtering effect}},\ }\href
  {https://doi.org/10.1088/1468-6996/9/1/014106} {\bibfield  {journal}
  {\bibinfo  {journal} {Sci. Technol. Adv. Mat.}\ }\textbf {\bibinfo {volume}
  {9}},\ \bibinfo {pages} {014106} (\bibinfo {year} {2008})}\BibitemShut
  {NoStop}%
\bibitem [{\citenamefont {Kyritsakis}\ \emph {et~al.}(2019)\citenamefont
  {Kyritsakis}, \citenamefont {Baibuz}, \citenamefont {Jansson},\ and\
  \citenamefont {Djurabekova}}]{Kyritsakis2019}%
  \BibitemOpen
  \bibfield  {author} {\bibinfo {author} {\bibfnamefont {A.}~\bibnamefont
  {Kyritsakis}}, \bibinfo {author} {\bibfnamefont {E.}~\bibnamefont {Baibuz}},
  \bibinfo {author} {\bibfnamefont {V.}~\bibnamefont {Jansson}},\ and\ \bibinfo
  {author} {\bibfnamefont {F.}~\bibnamefont {Djurabekova}},\ }\bibfield
  {title} {\bibinfo {title} {{Atomistic behavior of metal surfaces under high
  electric fields}},\ }\href {https://doi.org/10.1103/PhysRevB.99.205418}
  {\bibfield  {journal} {\bibinfo  {journal} {Phys. Rev. B}\ }\textbf {\bibinfo
  {volume} {99}},\ \bibinfo {pages} {205418} (\bibinfo {year}
  {2019})}\BibitemShut {NoStop}%
\bibitem [{\citenamefont {Zhang}(1999)}]{Zhang1999}%
  \BibitemOpen
  \bibfield  {author} {\bibinfo {author} {\bibfnamefont {S.}~\bibnamefont
  {Zhang}},\ }\bibfield  {title} {\bibinfo {title} {Spin-dependent surface
  screening in ferromagnets and magnetic tunnel junctions},\ }\href
  {https://doi.org/10.1103/PhysRevLett.83.640} {\bibfield  {journal} {\bibinfo
  {journal} {Phys. Rev. Lett.}\ }\textbf {\bibinfo {volume} {83}},\ \bibinfo
  {pages} {640} (\bibinfo {year} {1999})}\BibitemShut {NoStop}%
\end{thebibliography}

%
\end{document}